\documentclass{article}
\usepackage{graphicx}
\begin{document}                                 
\noindent {\bf Parametrization of Crab pulsar spectrum }
\vskip1.5cm
\noindent {\bf Ashok Razdan}

\noindent {\bf Astrophysical Sciences Division }

\noindent {\bf Bhabha Atomic Research Centre }

\noindent {\bf Trombay, Mumbai- 400085 }
\vskip 1.5cm
%\noindent {\bf Abstract :}

The recent detection of pulsed $\gamma$-ray from crab pulsar [1] by VERITAS $\gamma$-ray telescope
above 100 GeV can not be explained by standard pulsar models and data has been parametrized by
broken power law and power law with exponential cutoff.In this letter we explore the possibility of
using non extensive exponential function to parameterize the crab pulsar spectrum.  
Standard Statistical Physics ( Boltzmann-Gbibbs thermostatistics) holds as long as thermodynamic
extensivity  (additivity) holds i.e. when
(a) effective microcopic interactions are short range and
(b) systems evolve in
Euclidean like space-time ( a continuous and suffciently differentiable)
For two systems A and B entropy is additive
\begin{equation}
 S(A+B)= S(A)+S(B)
\end{equation}
Boltzmann-Gibbs(BG)entropy is additive and extensive.
BG approach fails 
(a) in systems with long range forces or long memory effects
(b) or  if systems evolve in non Euclidean space-time( i.e. fractals or multifractals).

Such systems which do not follow Boltzmann-Gibbs approach 
are called as non-extensive systems [2 and references therein].
For two systems A and B in non extenive approach
\begin{equation}
 S(A+B)= S(A)+S(B)+(1-q)S(A) S(B)
\end{equation}
where q is non extensive index.
Non-extensive statistics is based on two postulates of entropy and internal energy.
Non-extensive entropy is given as
\begin{equation}
S_q= k \frac{1}{q-1} (1- \int p(x,t)^q dx)
\end{equation}
Non-extensive entropy {11] is defined as
\begin{equation}
S_q  = \frac{ 1- \sum_i P_i^{q}}{q -1}
\end{equation}
and internal energy is 
\begin{equation}
U_q = \int  p_{i} ^q E_{i}= Tr \rho^q H 
\end{equation}
where $E_i$ is the energy spectrum i.e. the set of eigenvalues of the Hamiltonian H.
In the limit of q $\rightarrow$ 0 , entropy is given as
\begin{equation}
S= -k  p_{i} ln {p_i}
\end{equation}
which is Boltzmann-Gibbs Shannon entropy.
In non-extensive approach exponential is written as
\begin{equation}
 e_q ^x  = [1+ (1-q)x]^ \frac{1}{1-q}
\end{equation}
if $(1+(1-q)x) >$ 0 , otherwise $e_q ^x $=0.
Again  $ e_q ^{-x} =[1-(1-q)x]^ \frac{1}{1-q}$ holds.
We have used b*$ e_q ^{-cx} =b*[1-(1-q)cx]^ \frac{1}{1-q}$  to fit the data, where b and c are constants.
In our parameterization c=0.0003 and b=0.0002.
\begin{figure}
\includegraphics[angle=270,width=7.cm]{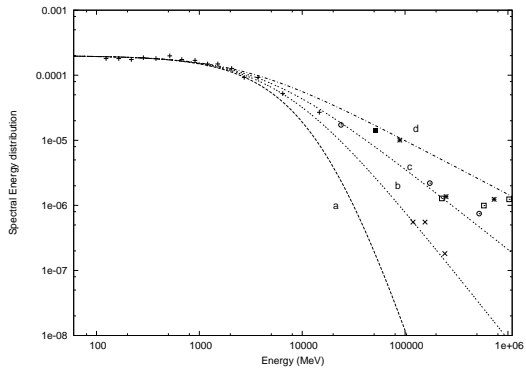}
\caption{ Parameterization of spectral energy distribution  is shown of crab pulsar data. In this figure hollow sphere show whipple data [9],+ corresponds to the
Fermi [3] data, filled square to CELESTE data [6],
Cross (X) to the VERITAS data [1] , * to  MAGIC [5] data, and circle corresponds
to MAGIC [4], STACEE [7] and HEGRA [8] data.
The fit corresponds to Non extensive exponential for for q=1.2 (curve a), q= 1.5 (curve b), q=1.8 (curve c) and
q=2.2 (curve d) respectively.}
\end{figure}

\end{document}